\documentclass[journal,letterpaper]{IEEEtran_nssmic}

\usepackage{cite}      

\usepackage{graphicx}  
\begin{document}
%
\title{Front-end electronics for the ALICE dimuon trigger RPCs}
%
%
\author{Philippe~Rosnet and Laurent~Royer, for the ALICE Collaboration
\thanks{Talk given by Ph.~Rosnet, LPC Clermont-Ferrand, CNRS/IN2P3 and Universit\'e Blaise Pascal, 63177 Aubi\`ere Cedex, France (Email: rosnet@in2p3.fr).}}
\maketitle

\begin{abstract}
A dedicated front-end electronics has been developed for the trigger
chambers of the ALICE muon spectrometer under construction at the future
LHC at CERN.
These trigger chambers are based on RPCs (Resistive Plate Chambers)
working in streamer mode.
The number of electronics channels (about 21000) and the fact that 
RPC pulsed signals have specific characteristics have led to the design
of a 8 channels front-end ASIC using a new discrimination technique.
The principle of the ASIC is described and the radiation hardness is 
discussed.
Special emphasis is put on production characteristics of about 4000
chips.
\end{abstract}


%
%
%
%
%

\section{Introduction to physics challenge}

ALICE~\cite{ALICE_TP} (A Large Ion Collider Experiment) is 
a detector designed for the study of nucleus-nucleus collisions 
at the future LHC (Large Hadron Collider).
Its physics program will address question concerning QCD (Quantum 
ChromoDynamics) of hot and dense nuclear matter produced in central
heavy ion collisions at a center of mass energy of 5.5 TeV.
The main goal of ALICE is to characterize a deconfined state of matter 
called the Quark-Gluon Plasma (QGP)~\cite{ALICE_PPR1}.
One of its most promising probes is the production of quarkonia 
(J/$\Psi$ or $\Upsilon$), which is expected to be decreased by color screening 
in the QGP~\cite{Vogt}.

The role of the ALICE forward spectrometer~\cite{ALICE_TDRmuon} is to 
reconstruct quarkonia in their dimuon channel in an angular acceptance 
of $[2^0,9^0]$.
To reach this physics goal, the spectrometer consists of a front and
small angle (beam shielding) absorber, a set of high resolution tracking 
chambers, a dipole magnet, an iron wall (muon filter) 
and a trigger system.

\section{Muon trigger system}

The trigger system is composed by two stations of two detector planes 
of about $6 \times 6$~m$^2$ separated by one meter.
The detectors are RPCs (Resistive Plate Chambers)~\cite{Santonico} 
working in streamer mode at a high voltage value of about 
8~kV~\cite{RPC_streamer}.
The necessary granularity implies to use strips of about 1, 2 and 4~cm
width projective to the interaction point in each of the four planes, 
for a total of twelve different strip widths.
Two perpendicular strip planes are used on each RPC (one on each side
of the gas gap) to allow a three dimensional hits reconstruction.
The total number of channels is then close to 21000. 

The muon trigger is involved in the level 0 (L0) of the general ALICE trigger 
system~\cite{ALICE_TDRtrigger}.
The timing constraint required by the L0 is a muon trigger signal delivered 
each 25~ns (40~MHz) to the central trigger processor less than 800~ns 
after the collision.
The architecture of the muon trigger electronics consists of the front-end
electronics which picks-up and processes the RPC pulses and sends logical 
signals to the local trigger electronics (234 boards). 
The role of the local trigger is to store all signals (in a sequence
of bit-patterns) in a pipeline memory and to identify single tracks
with a transverse momentum above pre-defined cuts (by using a
look-up-table) by means of a dedicated algorithm located in
FPGAs (Field Programmable Gate Arrays).
The regional (16 boards) and next the global (1 board) triggers collect
the information from the local boards in order to select single or dimuon
events from the full system.

\section{Front-end electronics principle}

\begin{figure}
\centering
\includegraphics[width=2.5in]{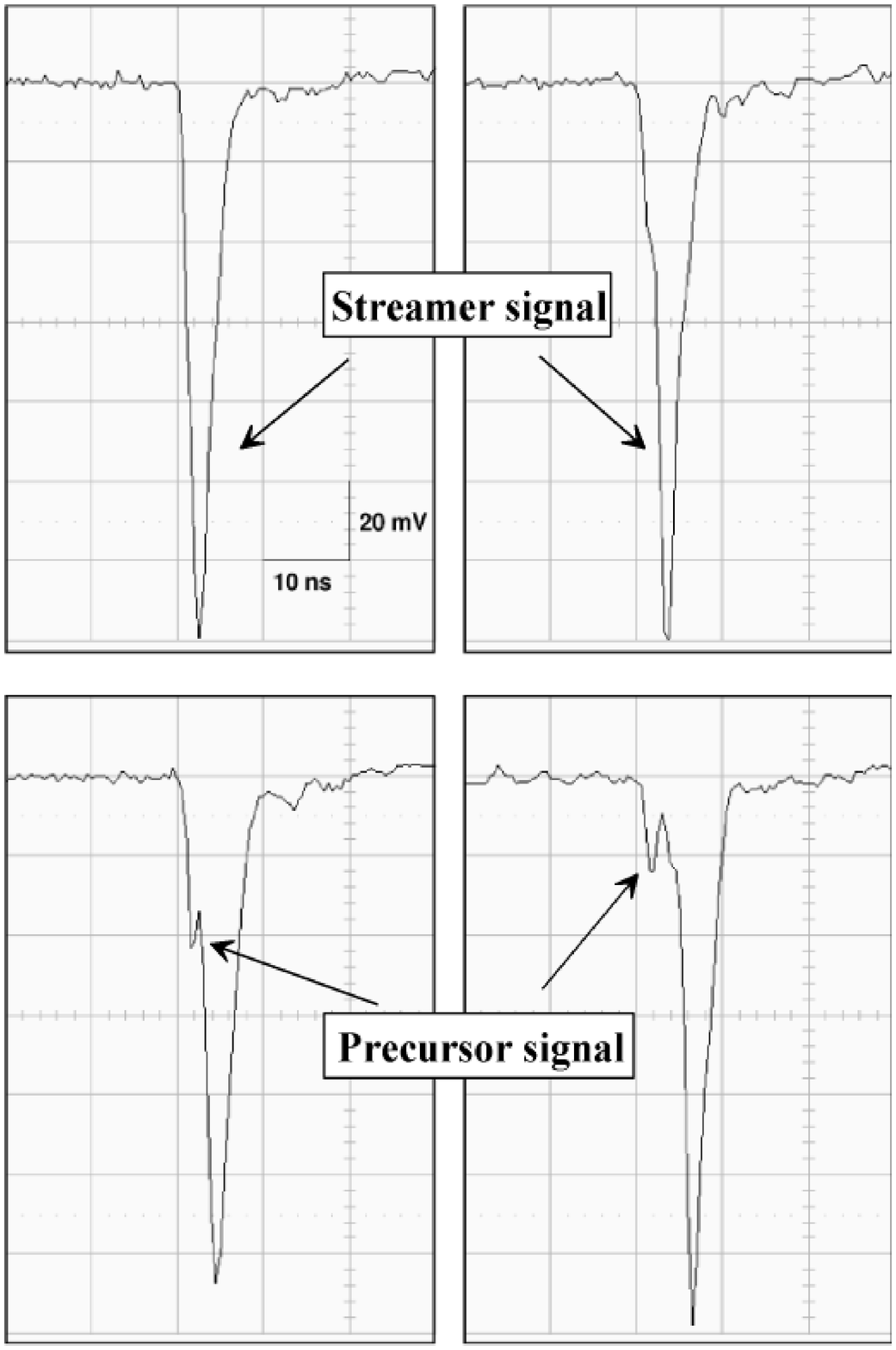}
\caption{Typical pulses picked-up on a single gas gap RPC operating
in streamer mode, with a digital scope (1 GHz bandwidth) via a short 
BNC cable (50 $\Omega$ impedance) synchronized with particles crossing
scintillator plates readout by photo-multipliers.}
\label{fig-RPC_pulses}
\end{figure}

As shown in Fig.~\ref{fig-RPC_pulses}, the RPC pulses are composed of
two peaks, called precursor and streamer.
The precursor has the particularity to be synchronous with the particle
crossing the RPC.
The second peak is characterized by a variable time jitter with respect to 
the precursor, but with a large amplitude as expected for a streamer pulse.

\subsection{ASIC main characteristics}

To take advantage of these properties a new discrimination technique,
called ADULT~\cite{FEE_ADULT} (A DUaL Threshold), has been developed 
and implemented in a 8 channels ASIC using the AustriaMicroSystems 
BiCMOS 0.8~$\mu$m technology~\cite{FEE_ASIC}.
This bipolar technology, with very low offset, is well adapted to trigger
on small amplitude signals.
The necessary number of ASIC for equipping all the muon trigger chambers 
is 2624 (not including spares).

\begin{figure*}
\centering
\includegraphics[width=7.0in]{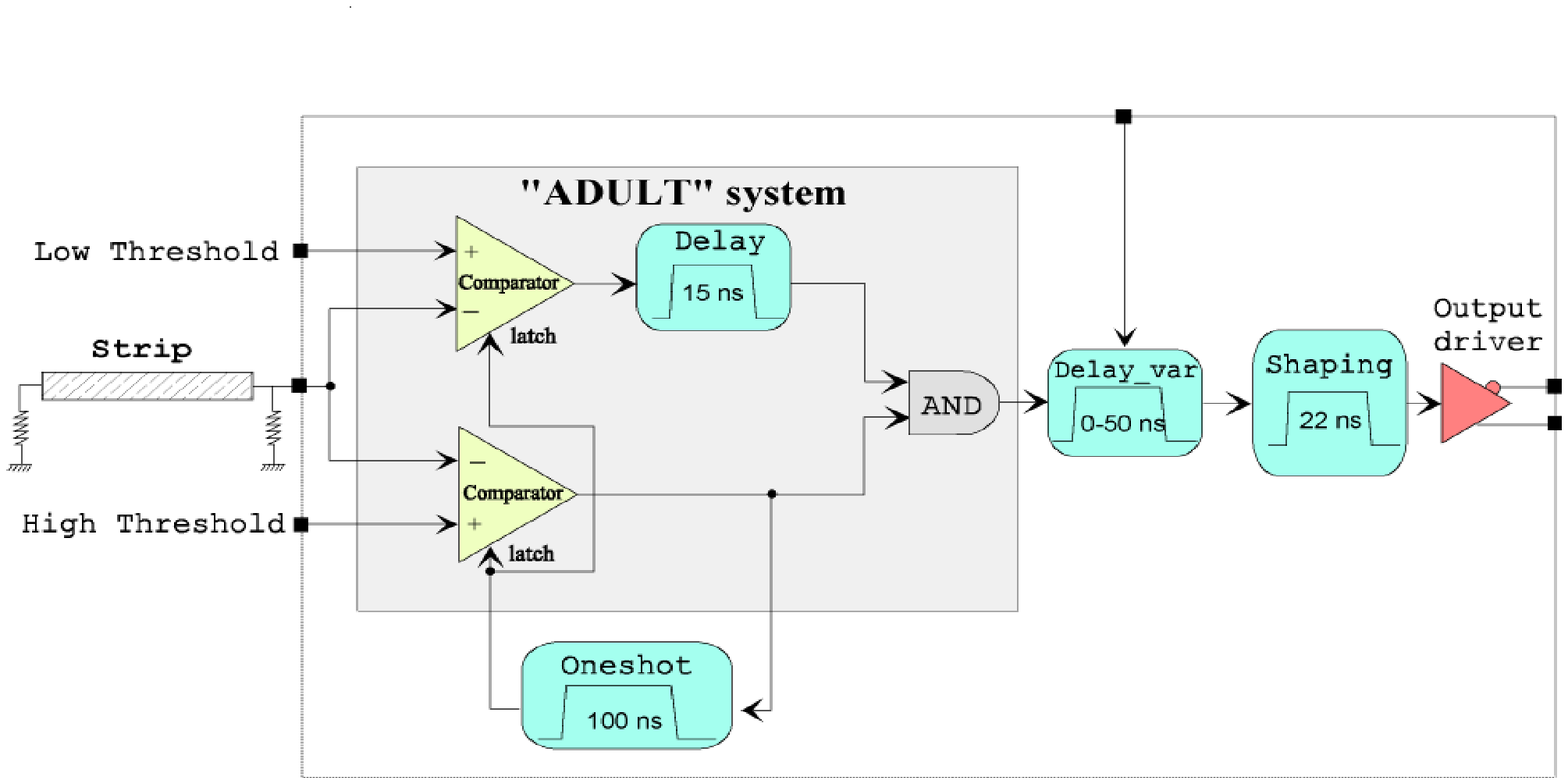}
\caption{Block diagram of a single electronics channel.}
\label{fig-ASIC_principle}
\end{figure*}

The schematic of a single electronics channel is shown in 
Fig.~\ref{fig-ASIC_principle}.
The precursor detection with a low threshold (typically 10~mV) provides
a good time reference, and the streamer validation with a high threshold
(about 100~mV) takes advantage of this RPC working mode: a large signal/noise 
ratio and a small cluster size (defined as the mean number of adjacent
strips fired).
A coincidence of these two discriminator output signals defines a hit of
a strip.
The time resolution obtained with this technique is comparable to the
avalanche, which is about 1~ns as compared to 3~ns typically obtained 
with a single discriminator in streamer mode.
The delayed (15~ns) low threshold comparator defines the reference
time by using the precursor as long as the streamer jitter is less 
than about 15~ns.
The oneshot function is used to latch the two comparators during 100~ns,
via a monostable, when a streamer signal has been validated to avoid 
re-triggering.
A remote control delay, up to 50~ns, common to the 8 channels of the
ASIC is tuned by an external DC voltage and allows to adjust the timing 
of the ASIC.
The signal is then converted into a (22-23)~ns logical LVDS signal in order 
to drive the signal through a twisted pair cable from the RPCs to the local 
electronics trigger boards.  

\begin{figure}
\centering
\includegraphics[width=2.5in]{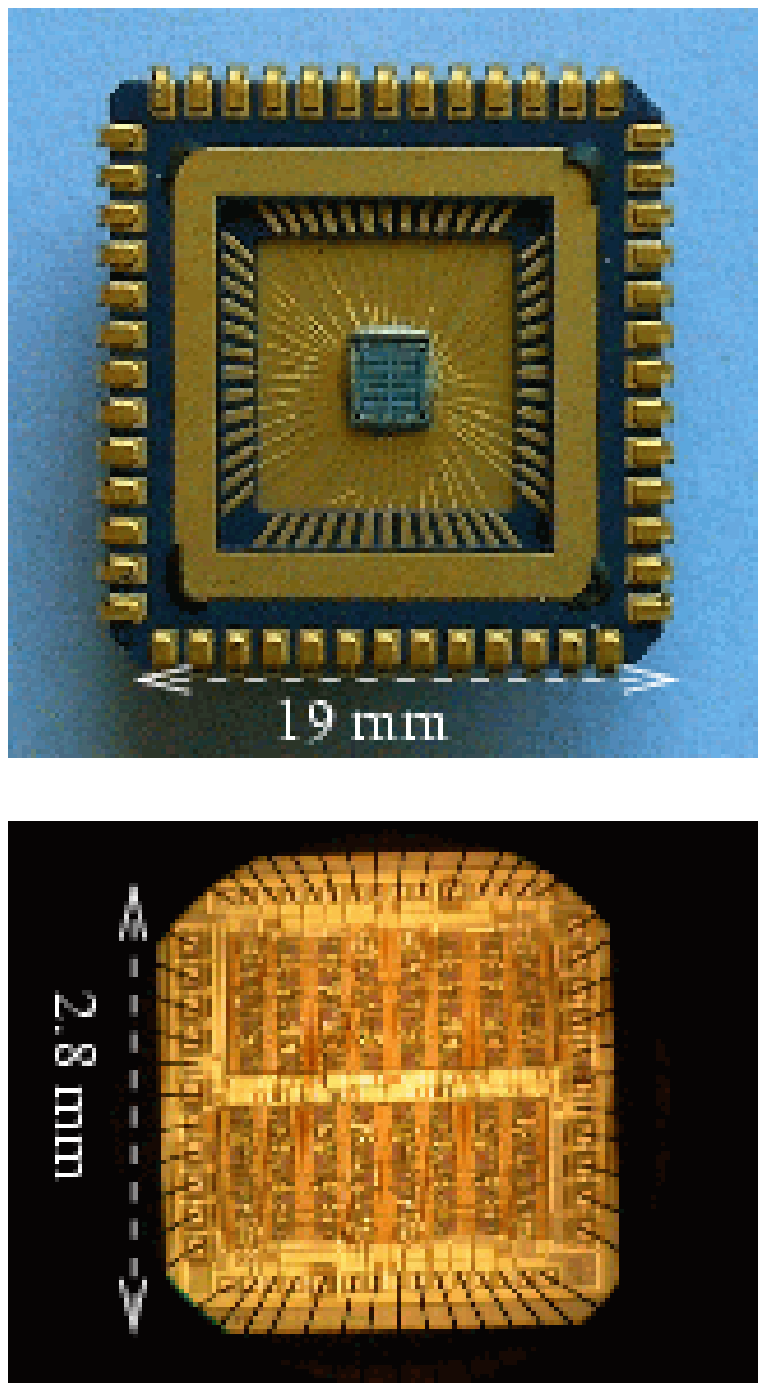}
\caption{ASIC after packaging (upper picture) and view of the silicon part
with a microscope (lower picture).}
\label{fig-ASIC_pictures}
\end{figure}

The main characteristics of an ADULT ASIC, as shown in
Fig.~\ref{fig-ASIC_pictures}, are the following:
\begin{itemize}
\item
AustriaMicroSystems BiCMOS 0.8~$\mu$m technology,
\item
8 electronics channels,
\item
die surface equal to about 8~mm$^2$,
\item
plastic packaging type PLCC 52 pins,
\item
power consumption: 10~mW/channel for $-2$~V and 80~mW/channel
for $+3.5$~V.
\end{itemize}

\subsection{Front-end boards supporting the ASIC}

\begin{figure*}
\centering
\includegraphics[width=6.0in]{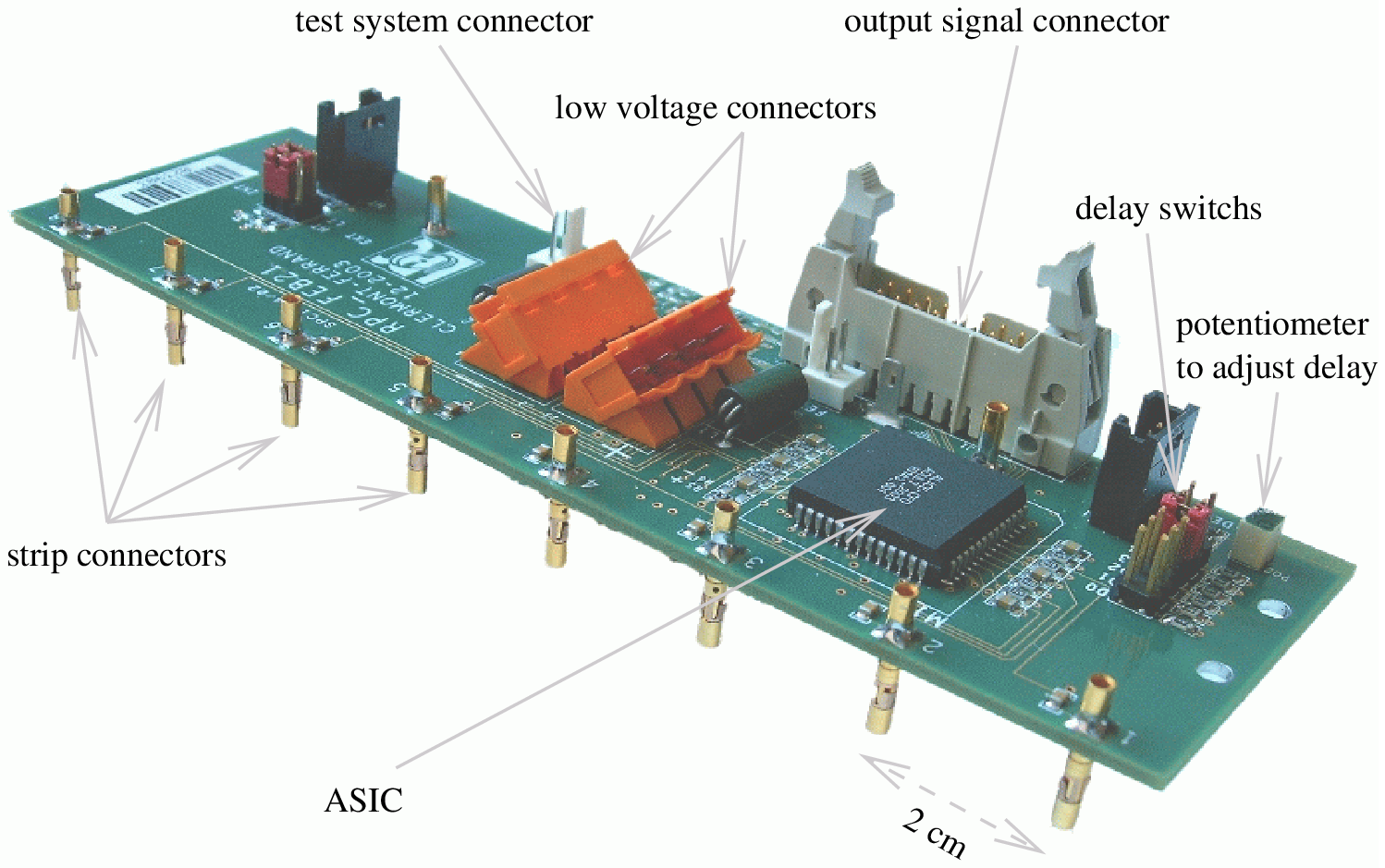}
\caption{Picture of a front-end board corresponding to 2~cm wide strips.}
\label{fig-FEB_picture}
\end{figure*}

The ASIC is implemented on a dedicated board developed to pick-up
the signals provided by the RPC strips, as shown in 
Fig.~\ref{fig-FEB_picture}.
Twelve different strip widths are necessary to equip the full detector, 
but only six boards with different mechanical characteristics are needed.
Two ASICs are used on the boards corresponding to strips of 1~cm width 
(16 channels) while the boards associated with strips of width 2~cm or 4~cm
(8 channels) contain only one ASIC.
Each single gas gap RPC provides positive pulses on one side and negative
pulses on the other side of the gas gap.
Then, the two polarities are implemented at the cabling level on 
the front-end boards.
Furthermore, the different distances between the boards and the trigger 
electronics located in racks above the chambers (6 meters high) lead
to different output cable length.
These differences were compensated for on the boards by implementing
delays in step of 7.5~ns corresponding to 1.5~m of cable.
To summarize, the total number of different boards after cabling is equal
to ten and each board can be configured with five possible delays, 
but the same ASIC is used on all boards.

In addition, a test system has been implemented on each front-end board:
a LVDS signal is received on the board, translated in TTL which,
and inverted, if appropriate, following the polarity associated to the board.
Then, a buffer allows to send an analogue pulse to each of the 8 ASIC
channels, simulating a RPC pulse with a width of 20~ns.

An adaptation board is associated to each front-end board at the other 
extremity of the strips.
It is a simple 50~$\Omega$ resistor which avoids signal reflection along 
the strips.

\section{Radiation tests}

Tests have been performed to check the radiation hardness in conditions 
close (or worse) than the expected ALICE working environment:
\begin{itemize}
\item
Cumulative effects due to low energy neutrons and to ionizing particles
(such as photons) which damage progressively the electronics components.
The simulations indicates 2.6~Gy over 10 LHC years for the boards closest
to the beam.
\item
Single event effects due to energy hadrons ($> 30$~MeV) which induce 
malfunctioning or may even destroy components. 
The simulations indicates $2 \times 10^{11}$~hadrons/cm$^2$ over 10 LHC 
years (and integrated over the energy spectrum) for the boards closest
to the beam.
\end{itemize}
The first test was done at the Gamma Irradiation Facility (GIF) at CERN
providing $\gamma$ of 662~keV from a Cs source. 
The total dose received by the electronics was about 0.25~Gy.
A second test was performed at the neutron generator of the LPC 
Clermont-Ferrand which delivers neutrons of 14.1~MeV.
The total neutron fluence received by the electronics components 
was $3.8 \times 10^{10}$~n/cm$^2$ corresponding to a dose of 2.5~Gy 
which was measured with the help of PIN diodes.
The third test was done with protons of 60~MeV at the Paul Scherrer Institute
(PSI) at Z\"urich.
The fluence measured on the board was $2.3 \times 10^{11}$~p/cm$^2$.

\begin{figure}
\centering
\includegraphics[width=3.5in]{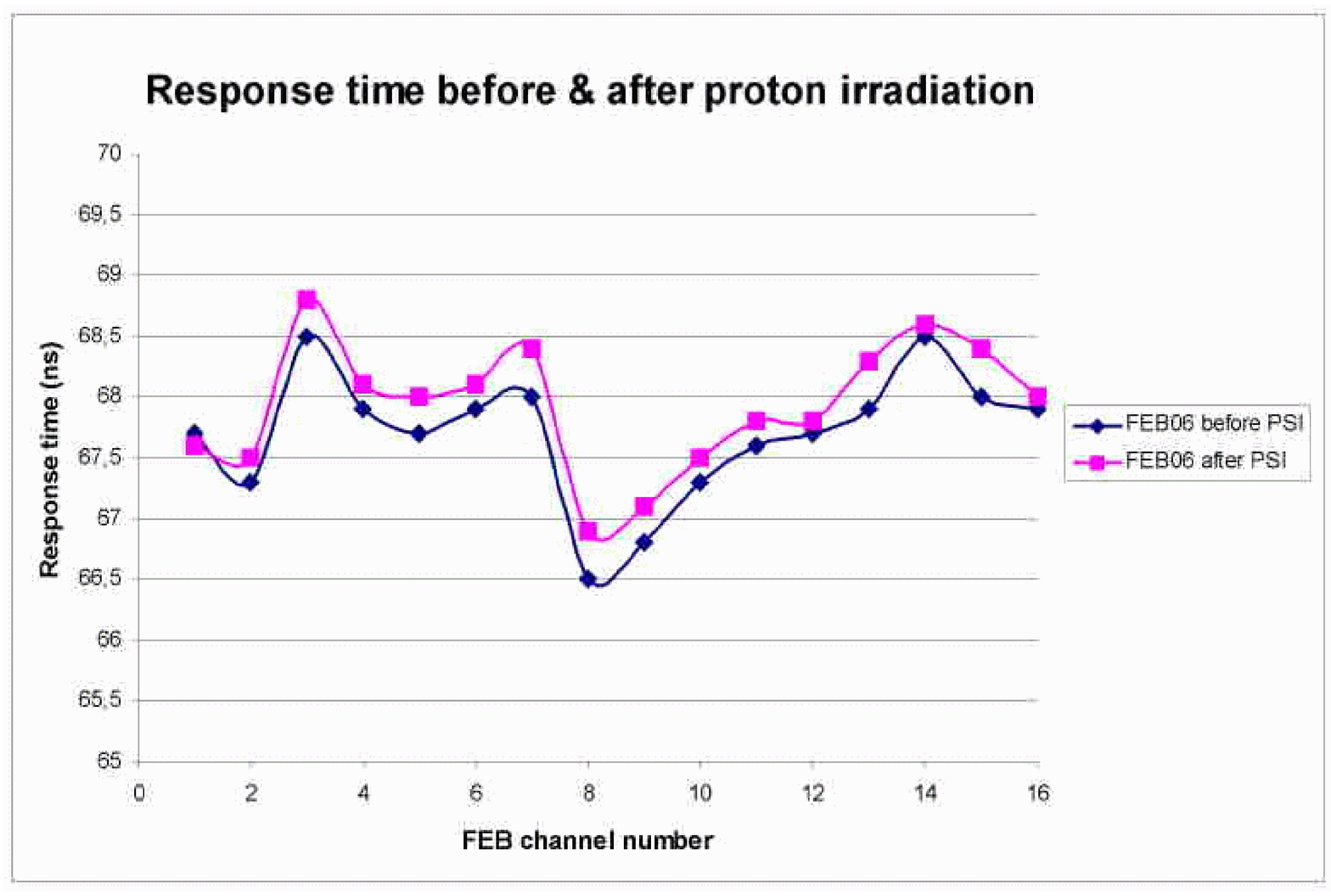}
\caption{Response time for 16 channels (2 ASICs) before and after irradiation
with the proton beam at PSI-Z\"urich.}
\label{fig-PSI_radiation}
\end{figure}

During all these tests, the front-end boards were active and 
pulsed signals from a generator (or the internal test system of the board) 
were sent (or activated) to check the ASIC (or the whole chain: test 
system plus ASIC) for each electronics channel.
For the three campaigns, no problem has been observed, as illustrated for 
example by Fig.~\ref{fig-PSI_radiation} showing the response time obtained 
with two ASICs of the same board before and after irradiation to protons
at PSI.

\section{Front-end electronics production}

The main constraint on this front-end electronics is to deliver 
the signal of any of the 21000 channels to the muon trigger electronics 
in a time window less than 25~ns (corresponding to the LHC clock).
By taking into account all the time dispersion sources coming mainly from
the RPC itself, the strip length (up to 72~cm), the output signal cables 
(up to 20 meters), the requirement for the front-end electronics is a time 
dispersion less than 4~ns.

To check this time requirements and others parameters of each electronics 
channels, the test of the production is divided in two steps:
\begin{itemize}
\item[(i)]
a working test and characterization of each ASIC (whole production done 
end of the year 2003 corresponding to 3880 chips),
\item[(ii)]
a tuning of the response time of the board (mean value) with the help 
of a potentiometer and a measurement of each parameters (about 12) 
of each electronics channel of the front-end boards (production done 
during summer 2004).
\end{itemize}

\subsection{Automatic test bench design}

The test bench is based on a special card equipped with relays 
which allow to switch to any electronics input and output channel 
of a board (the six different mechanical front-end boards can be handled).
The input signal is generated by a pulse generator simulating RPC signals, 
including precursor and streamer peaks, with the possibility to vary
each parameters of the signal (amplitude of each peak, time jitter between 
precursor and streamer, ...).
A copy of this signal is sent to a scope (500 MHz bandwidth) to measure
exactly its characteristics.
The scope is also used to measure the output LVDS signal generated by 
the front-end board.
The apparatus of the test bench are controlled by Labview via GPIB and 
the relays via a DIO card. 

\subsection{Test of the ASIC production}
 
\begin{figure}
\centering
\includegraphics[width=3.5in]{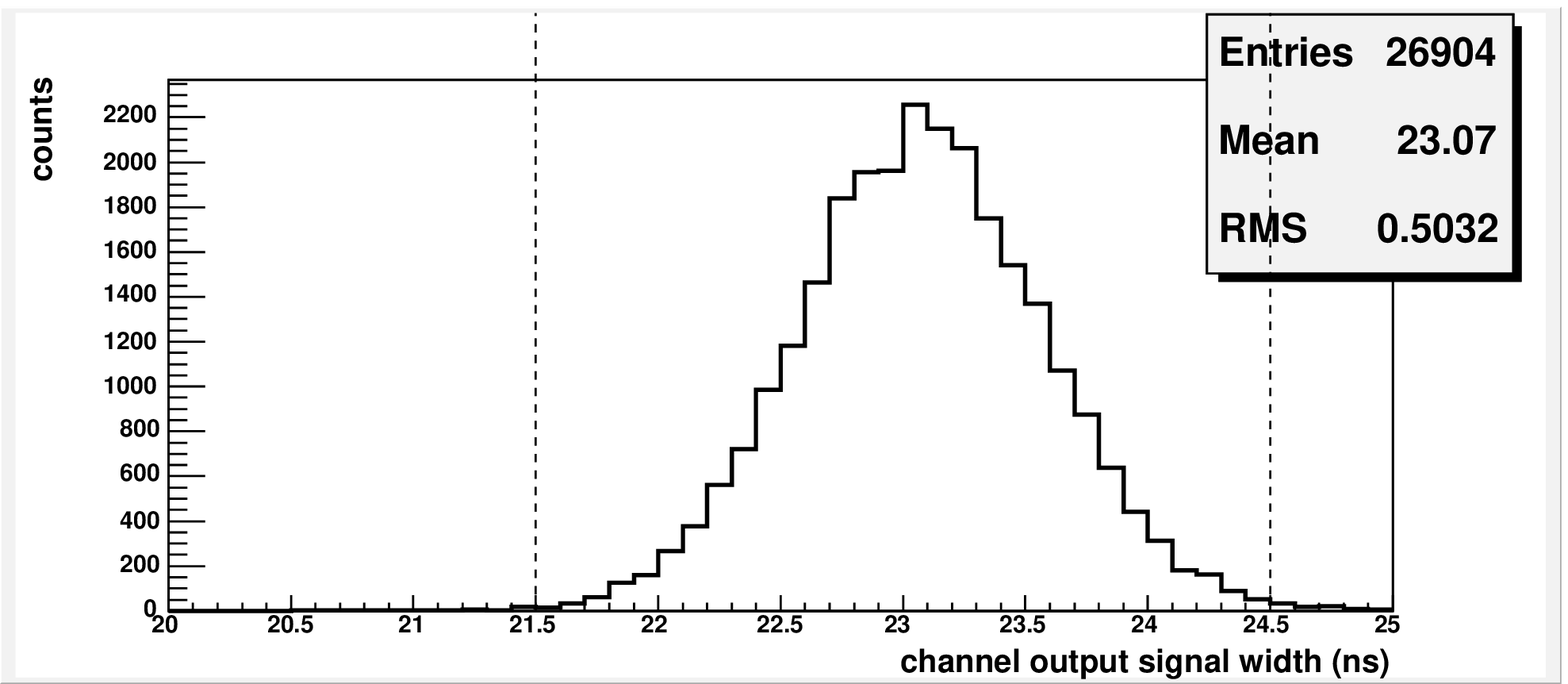}
\includegraphics[width=3.5in]{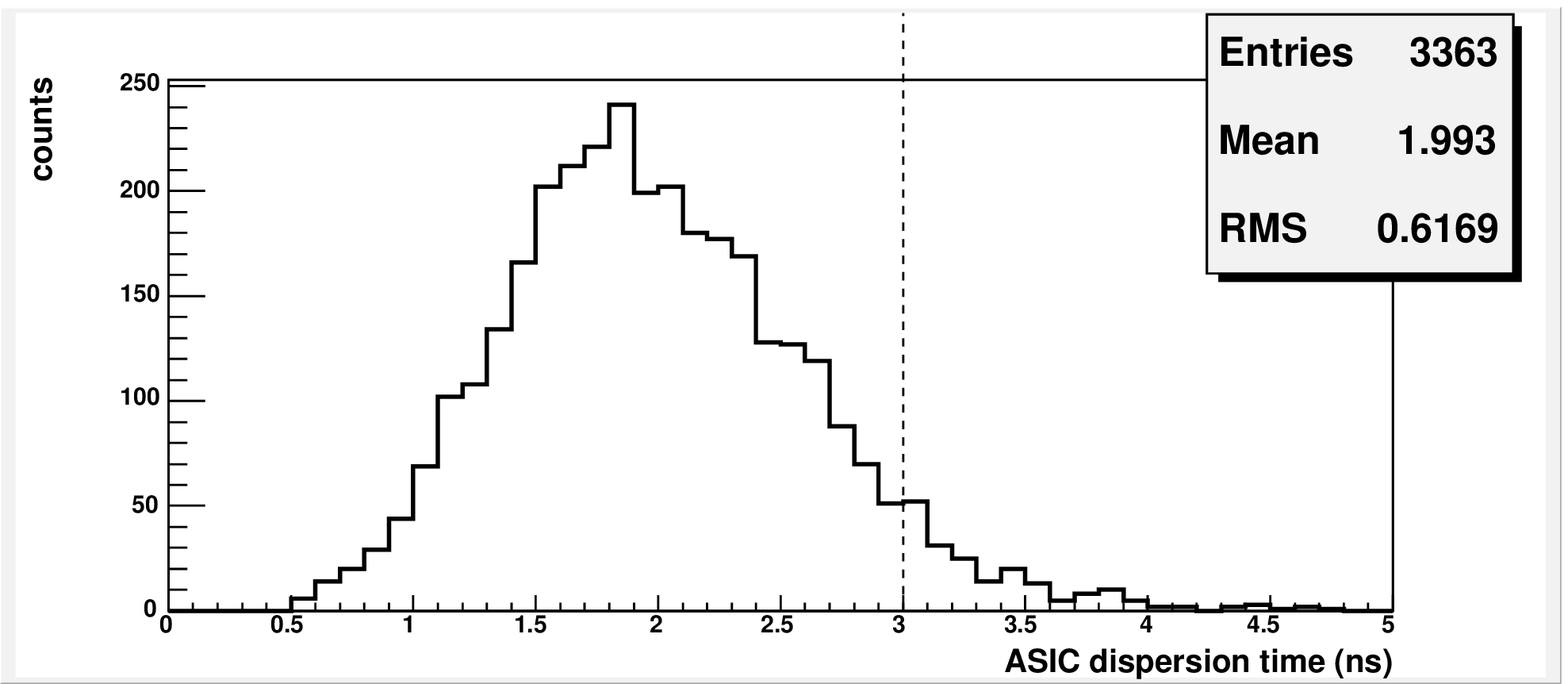}
\caption{ASIC production characteristics: channel output signal width
$w_{ch}$ (upper plot) and ASIC dispersion time $\Delta t_{ASIC}$ 
(lower plot).
The dotted lines on each plot represent the requirements.} 
\label{fig-ASIC_production}
\end{figure}

In practice, for the test of the ASICs, the same front-end board equipped 
with a socket allowing to plug and unplug easily each chip was used.
For each ASIC, the 8 output signals were characterized by measuring
the amplitude of the LVDS signal $A_{ch}$, its width $w_{ch}$,  
and the time difference between the slowest and the fastest 
channels $\Delta t_{ASIC}$ (which can be associated to the response
time dispersion of the ASIC).
Fig.~\ref{fig-ASIC_production} shows the channel output signal width
distribution and the ASIC time dispersion. 
The requirements for each parameter are: $A_{ch} = (800 \pm 100)$~mV,
$w_{ch} = (23.0 \pm 1.5)$~ns and $\Delta t_{ASIC} < 3$~ns.

Over 3880 ASICs, the results are the following:
\begin{itemize}
\item
5.7\% were not working (due to short-circuits, ...),
\item
11.8\% were working but with at least one parameter outside limits 
(about half due to output signal width and half due to response time 
dispersion),
\item
82.5\% were within specifications.
\end{itemize}
This means that 3280 ASICs are available for ALICE front-end board
production (while 2624 are needed).

\subsection{Test of the pre-production boards}

The whole production of the boards (including spares) was done during 
summer 2004. 
At the time of the writing, the tests have just started and are scheduled 
to finish in summer 2005.

\begin{figure}
\centering
\includegraphics[width=3.5in]{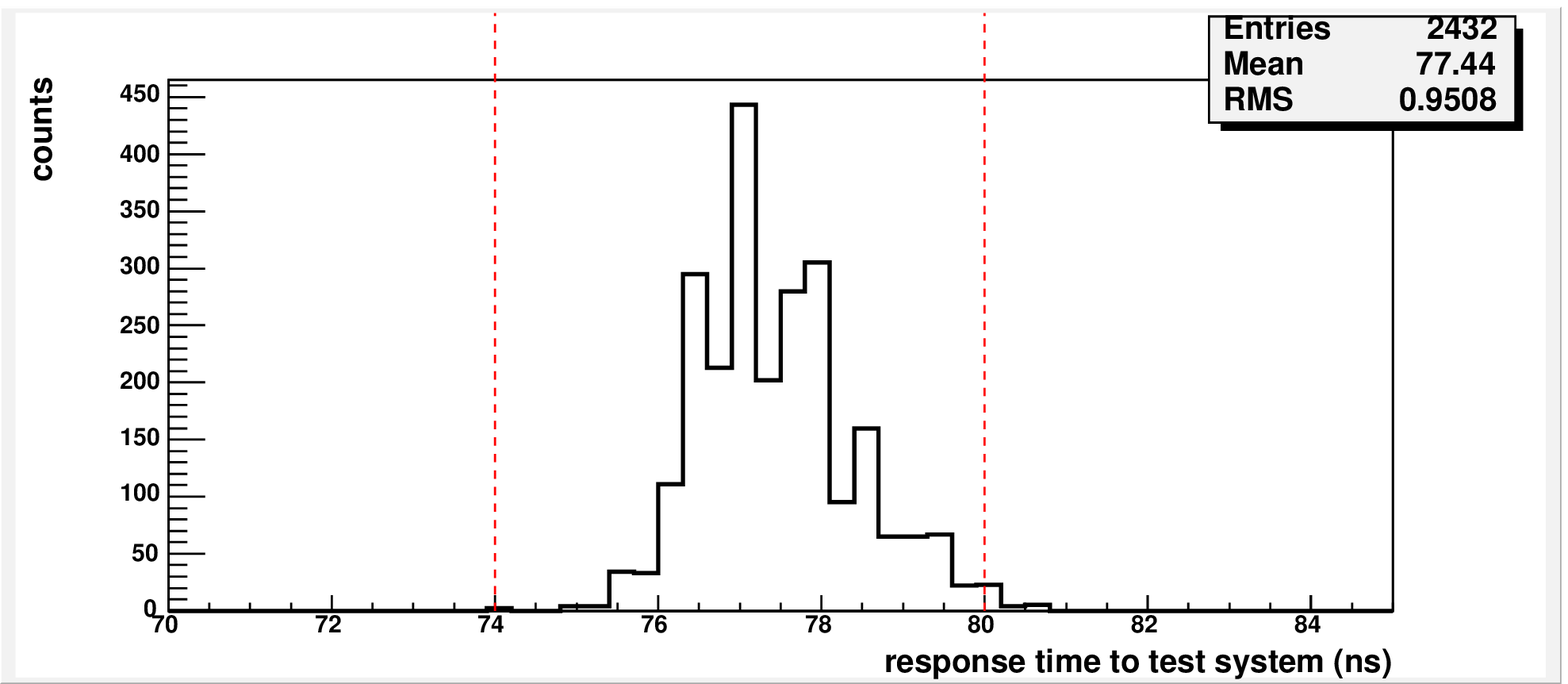}
\includegraphics[width=3.5in]{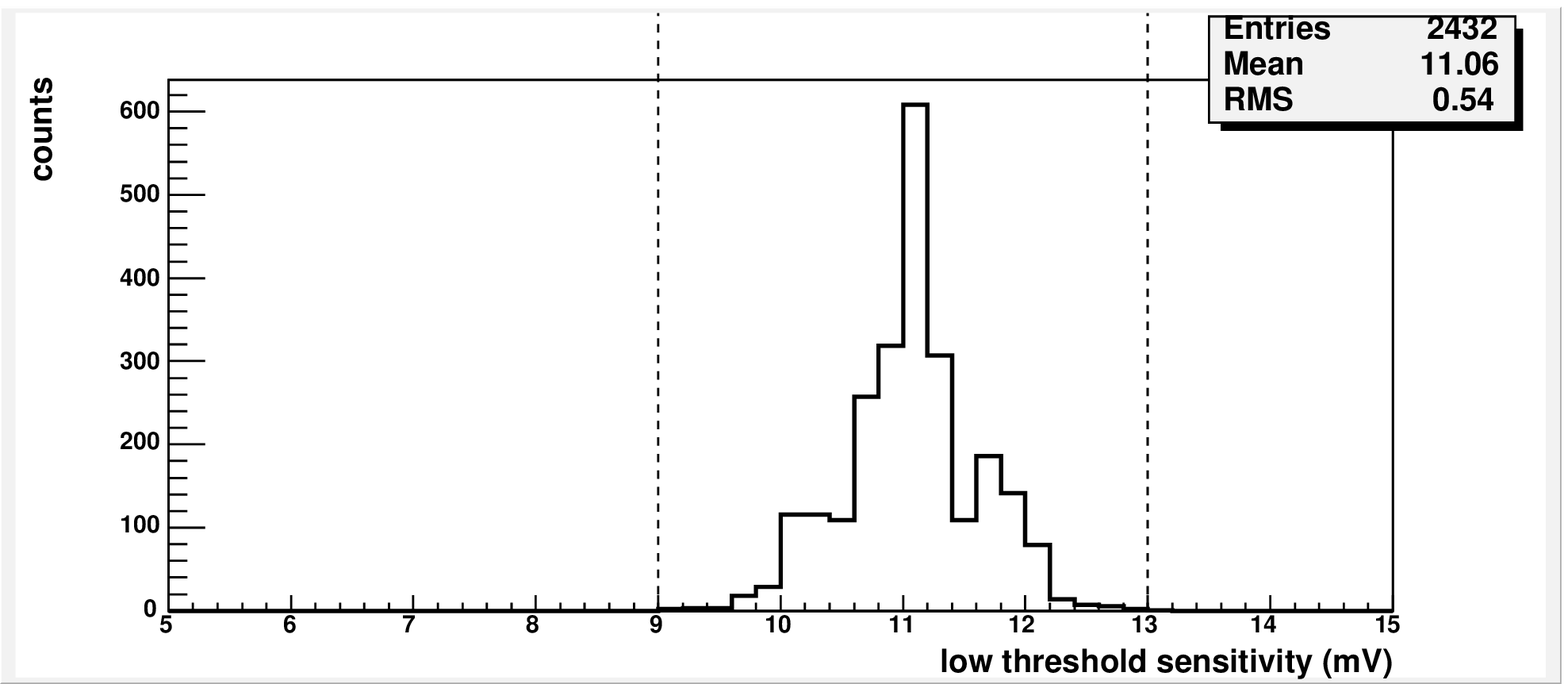}
\includegraphics[width=3.5in]{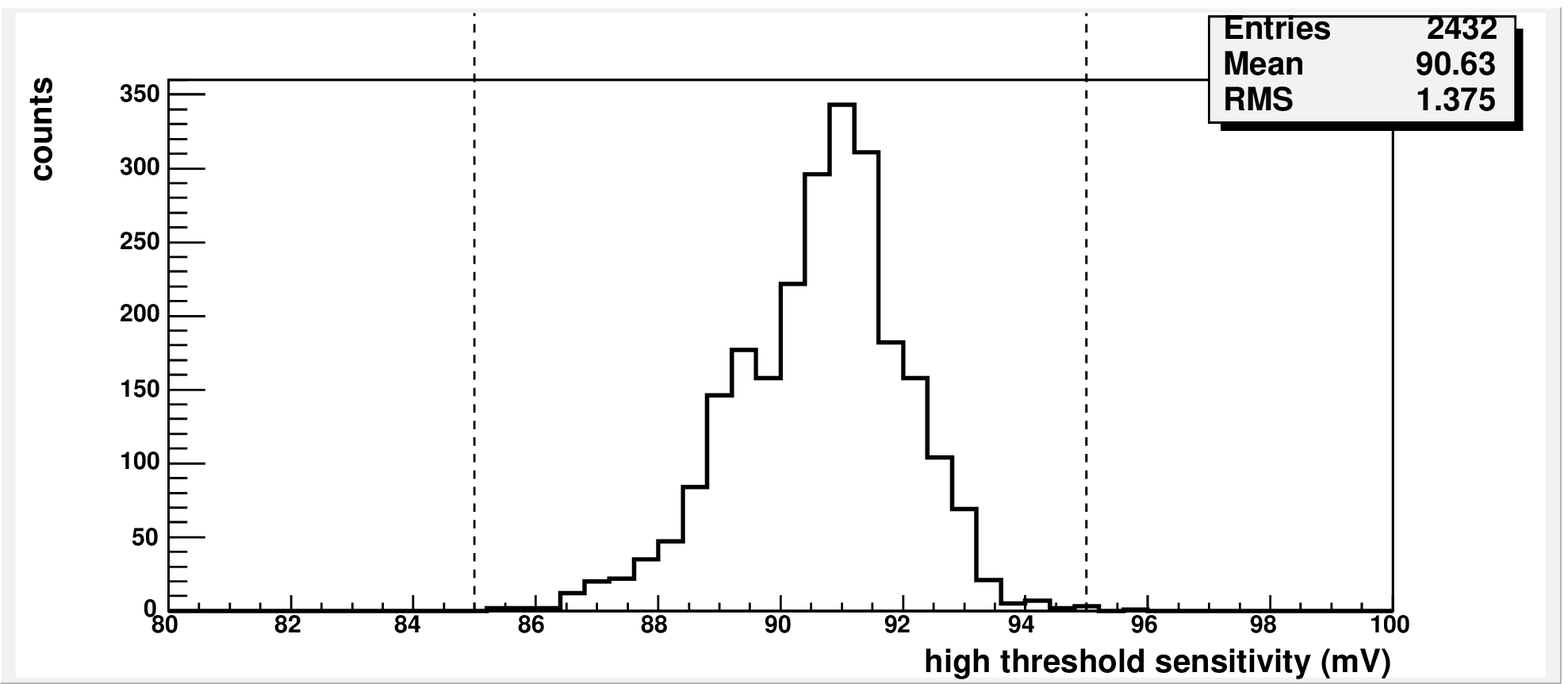}
\caption{Front-end boards pre-production characteristics: 
response time with the internal test system (upper plot),
low threshold discriminator (middle plot)
and high threshold discriminator (lower plot).
The dotted lines on each plot represent the requirements.}
\label{fig-FEB_production}
\end{figure}

However, the test of a pre-production representing about 12\% 
the whole production (286 boards) was performed at the beginning of 2004.
These boards are devoted to the test bench of the RPCs located at Torino.
Each board has been tested and all the parameters measured.
Fig.~\ref{fig-FEB_production} displays the results obtained for
three parameters: the response time to the internal test system,
the low and high threshold discriminator.
The most crucial parameter, {\it e.g.} the response time, shows 
good behavior (the RMS of the distribution is less than 1~ns) 
with respect to the ALICE requirement (all 21000 channels in 
a time window of 4~ns due to the front-end dispersion alone).
The value of the discriminator thresholds are relatively well peaked
around the expectations.

During the test, it appeared that 5 ASICs were not working over
316 (among the 286 boards, 30 have two ASICs), namely about 2\%.
The other problems met are attributed to the cabling and concerned about
4\% of the boards (only one problem on printed circuit was reported).
After intervention, only 2\% of the board have at least one channel
with one or more parameters outside specifications.
In this case, the only possibility to solve the problem was to change
the ASIC.

\section{Conclusion}

The dedicated front-end electronics of the ALICE muon trigger, for
RPC in streamer mode, is based on a 8 channels ASIC which has been 
produced and tested.
Its performance fulfils the ALICE requirements, both in terms of timing 
and concerning the low threshold discriminator sensitivity needed 
for the ADULT technique.
The yield of bad chips (less than 6\%) is acceptable. 
The pre-production of front-end boards supporting the ASIC 
shows that it is possible to tune each board such that all channels
are within a time window of 4 ns.
The fraction of chips to be changed after the cabling phase is about 4\%,
leaving a sufficient number of spares for the lifetime of the ALICE
experiment.

\section*{Acknowledgment}

The authors would like to thank the R\'egion Auvergne (France) for 
funding support.

\end{document}